\documentclass{aa}  
\usepackage{graphicx}
\usepackage{txfonts}
\begin{document}
%
%
\newcommand{\tuc}{\rm $J$=1--0}            
\newcommand{\tdu}{\rm $J$=2--1}         
\newcommand{\ttd}{\rm $J$=3--2}         
\newcommand{\doce}{\rm $^{12}$CO}       
\newcommand{\trece}{\rm $^{13}$CO}      
\newcommand{\gsim}{\raisebox{-.4ex}{$\stackrel{>}{\scriptstyle \sim}$}}
\newcommand{\lsim}{\raisebox{-.4ex}{$\stackrel{<}{\scriptstyle \sim}$}}
\newcommand{\psim}{\raisebox{-.4ex}{$\stackrel{\propto}{\scriptstyle \sim}$}}
\newcommand{\kms}{\mbox{km~s$^{-1}$}}
\newcommand{\jyb}{\mbox{Jy~beam$^{-1}$}}
\newcommand{\s}{\mbox{$''$}}
\newcommand{\mloss}{\mbox{$\dot{M}$}}
\newcommand{\my}{\mbox{$M_{\odot}$~yr$^{-1}$}}
\newcommand{\ls}{\mbox{$L_{\odot}$}}
\newcommand{\ms}{\mbox{$M_{\odot}$}}
\newcommand{\mm}{\mbox{$\mu$m}}
\def\arcdeg{\hbox{$^\circ$}}
\newcommand{\secp}{\mbox{\rlap{.}$''$}}
\newcommand{\secs}{\mbox{\rlap{.}$^{\rm s}$}}
\newcommand{\um}{\mbox{$\mu$m}}
\newcommand{\h}{$^{\rm h}$}
\newcommand{\m}{$^{\rm m}$} 
\newcommand{\irc}{IRC\,+10420}         
\newcommand{\afg}{AFGL\,2343}         
\title{ The chemical composition of the circumstellar envelopes around 
yellow hypergiant stars \thanks{ Based on observations carried out
with the IRAM Pico Veleta 30m telescope. IRAM is supported by INSU/CNRS (France), MPG (Germany)
and IGN (Spain).}  }
\titlerunning{Chemistry in the CSEs around YHGs}
%
%
   \author{
          G. Quintana-Lacaci \inst{1}
          \and
          V. Bujarrabal \inst{1}
          \and
          A. Castro-Carrizo \inst{2}
	  \and
          J. Alcolea \inst{3}
          }
   \offprints{g.quintana@oan.es}
   \institute{
     Observatorio Astron\'omico Nacional (IGN), Apdo. 112, 
                                   E-28803 Alcal\'a de Henares, Spain \\
              \email{(g.quintana,v.bujarrabal)@oan.es}
         \and
     Institut de RadioAstronomie Millim\'etrique, 300 rue de la Piscine,  
                                   38406 Saint Martin d'H\`eres, France \\
              \email{ccarrizo@iram.fr}
	\and
	Observatorio Astron\'omico Nacional (IGN), Alfonso XII N$^{\b o}$3,
				E-28014 Madrid, Spain \\
               \email{j.alcolea@oan.es}
     }

\date{Received 2007 / Accepted }

 
  \abstract 
 {The yellow hypergiant stars (YHGs) are extremely luminous and massive objects whose
general properties are poorly known. Only two of this kind of star
show massive circumstellar envelopes, IRC\,+10420 and AFGL\,2343.}
  {We aim to study the chemistry of the circumstellar envelopes around these two sources, 
by comparison with well known AGB stars and 
protoplanetary nebulae. We also estimate the abundances of the observed 
molecular species.}
  {We have performed single-dish observations of different transitions for 
twelve molecular species. We have compared the ratio of the intensities of the 
molecular transitions and of the estimated abundances in AFGL\,2343 and IRC\,+10420 with
 those in O-rich and C-rich AGB stars and protoplanetary nebulae.}
  {Both YHGs, AFGL\,2343, and IRC\,+10420, have been found to have an O-rich chemistry
similar to that in O-rich AGB stars, though for AFGL\,2343 the emission of most molecules 
compared with \trece\ lines is relatively weak. Clear differences with the other evolved sources appear
when we compare the line intensity corrected for distance and the profile
widths which are, respectively, very intense and very wide in YHGs. The abundances 
obtained for IRC\,+10420 agree with those found in AGB stars, but in general those found in 
AFGL\,2343, except for $^{13}$CO, are too low. This apparently low molecular 
abundance in AFGL\,2343 could be due to the fact that these molecules are present only in an inner region of the shell where the mass is relatively low.}
  {}


   \maketitle
\section{Introduction}

The Yellow hypergiants stars (YHGs) are among the most luminous (5.3
$\leq$ log$L$[\ls] $\leq$ 5.9) and massive ($M_{\rm init} \sim$ 20 \ms) stars 
(see, as general references, de Jager 1998, Jones et al$.$ 1993, Humphreys 1991).
These objects are thought to be post-red supergiants evolving bluewards in the 
HR diagram, but the details of such an evolution are still unknown. In at least a few 
of them, the stellar temperature is rapidly increasing. For instance, the spectral 
type of the hypergiant IRC\,+10420 has changed from F8Ia to A5Ia in just 20 yr 
(Klochkova et al.\ 1997,{ Oudmaijer et al. 1996, Oudmaijer 1998). Humphreys et al.
(2002) showed, however, that the wind in this source is optically thick, suggesting 
that the apparent spectral type changes are due to variations in the wind rather 
than to interior evolution.}

Although it is thought that, during the red and yellow phases, 
these heavy stars eject as much as one half of their initial mass 
(e.g.\ Maeder \& Meynet 1988, de Jager 1998), only two YHGs, IRC\,+10420 (= IRAS\,19244+1115) 
and AFGL\,2343 (= IRAS\,19114+0002 = HD179821) 
are known to have very heavy circumstellar 
envelopes (CSEs). Those CSEs were detected in molecular line 
emission, dust-scattered light and IR emission (see Hawkins et al$.$ 1995, 
Meixner et al$.$ 1999, Bujarrabal et al$.$ 1992, { Humphreys et al$.$ 1997}, 
Neri et al$.$ 1998, Bujarrabal et al$.$ 2001, Castro-Carrizo et al.\ 2001). 

Recent results from Castro-Carrizo et al.\ (2007) show that these 
circumstellar envelopes have several solar masses, very high expansion 
velocities ($\sim$ 35 \kms), and that such an environment was formed in the last $\sim 6000$\,yr.
The envelope properties there found are compatible with a mass loss driven by 
radiation pressure. { However, in the case of IRC\,+10420
the existence of infalling material 
suggests that other processes are also present
in the inner parts of the envelope (Humphreys et al.\ 2002).} 
Other properties of these molecular shells, like their chemical composition, 
have not been well studied.

This paper is devoted to the study of the chemistry in the
envelopes around IRC\,+10420 and AFGL\,2343. We have observed rotational lines of
 several molecules and, in particular, we compare the line intensities and 
molecular abundances found in these objects with those of AGB CSEs
and protoplanetary nebulae (PPNe).

\begin{figure*}
\centering
 \includegraphics[angle=-90, width=17cm ]{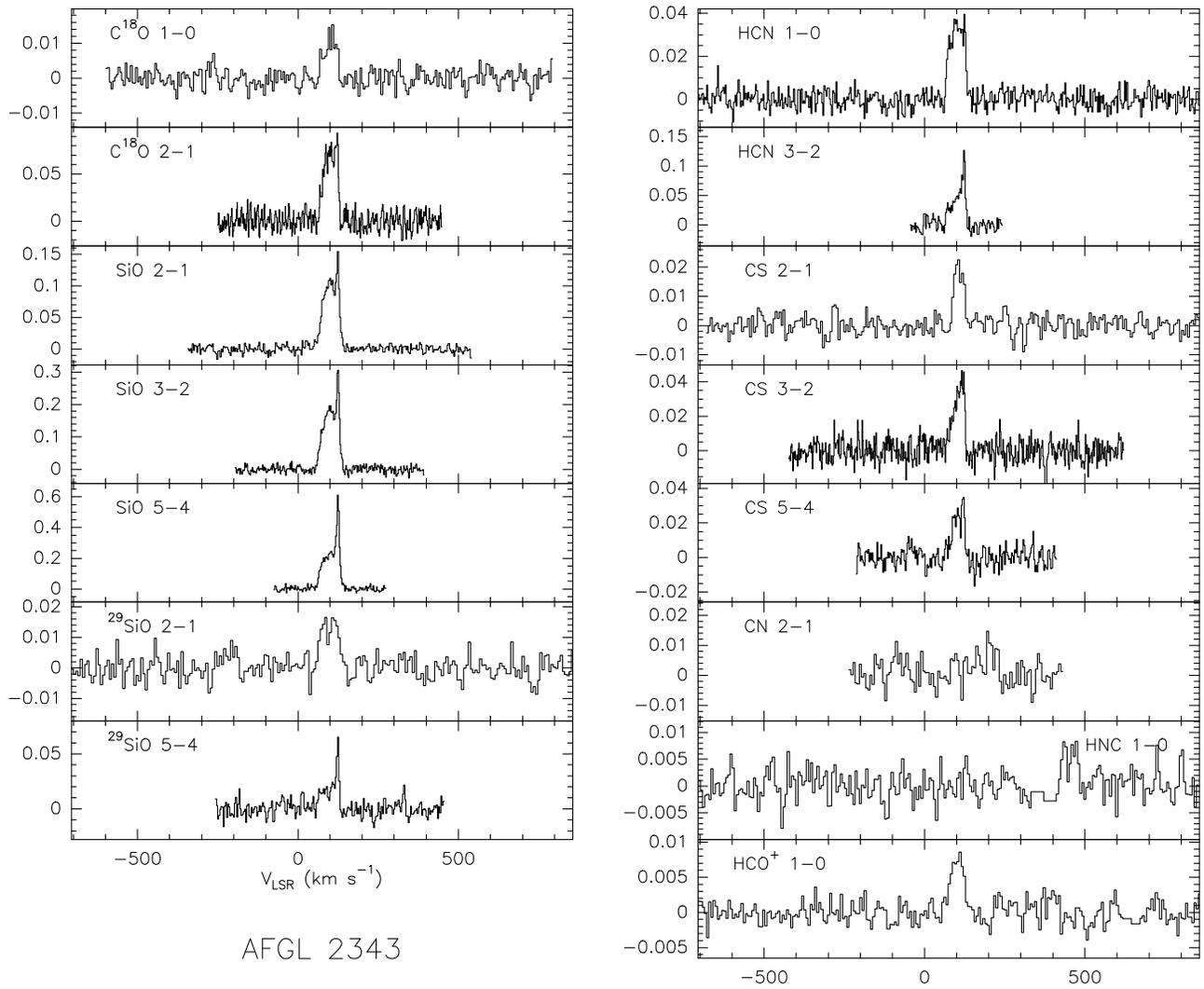}
 \caption{Observed spectra towards AFGL\,2343. The intensity scale is given in units of
 main-beam brightness temperature (K).}
\end{figure*}

Both IRC\,+10420 and AFGL\,2343 have very probably O-rich chemistry, in view of their OH maser emission (Likkel  1989, Reid et al.\
1979; among evolved objects, only O-rich stars are found to
emit in OH masers), silicate-rich circumstellar grains (e.g.\ 
Molster et al.\ 2002) and C-poor atmospheric composition (Klochkova et al.\ 1997, Th\'evenin et al.\ 2000). 

Castro-Carrizo et al.\ (2007) argued that AFGL\,2343 was a YHG rather than 
a PPN as proposed by Josselin \& L\`ebre (2001). 
As we will see, our molecular data confirm this, since the molecular line properties of these objects are quite similar and significantly different from those usual in both PPNe and AGB stars. { On the other hand, the nature of IRC\,+10420 is well known (see Jones et al. 1993, Oudmaijer et al. 1996)}. Our data will be used, in general, to determine the main chemical properties of the envelopes surrounding IRC\,+10420 and AFGL\,2343. 

\section{Observations}

We have used the IRAM 30m telescope, at Pico Veleta (Spain), to observe mm-wave molecular lines in the yellow hypergiants IRC\,+10420 and AFGL\,2343. The observed lines are:

\begin{itemize}
\item In the 3mm band: C$^{18}$O $J$=1--0, HCN $J$=1--0, H$^{13}$CN $J$=1--0, SiO 
$J$=2--1, $^{29}$SiO $J$=2--1, CS $J$=2--1, CN $N$=1--0, SiS $J$=5--4, SO $J_K$=$2_2$--$1_1$, HC$_3$N $J$=10--9, HNC $J$=1--0, and HCO$^+$ $J$=1--0.

\item In the 2mm band: SiO $J$=3--2 and CS $J$=3--2.

\item In the 1mm band: C$^{18}$O $J$=2--1, HCN $J$=3--2, SiO $J$=5--4, $^{29}$SiO
$J$=5--4, CS $J$=5--4, CN $N$=2--1, SiS $J$=15--14, HNC $J$=3--2, and HCO$^+$ $J$=3--2.
\end{itemize}

\begin{figure*}
\centering
 \includegraphics[angle=-90, width=17cm ]{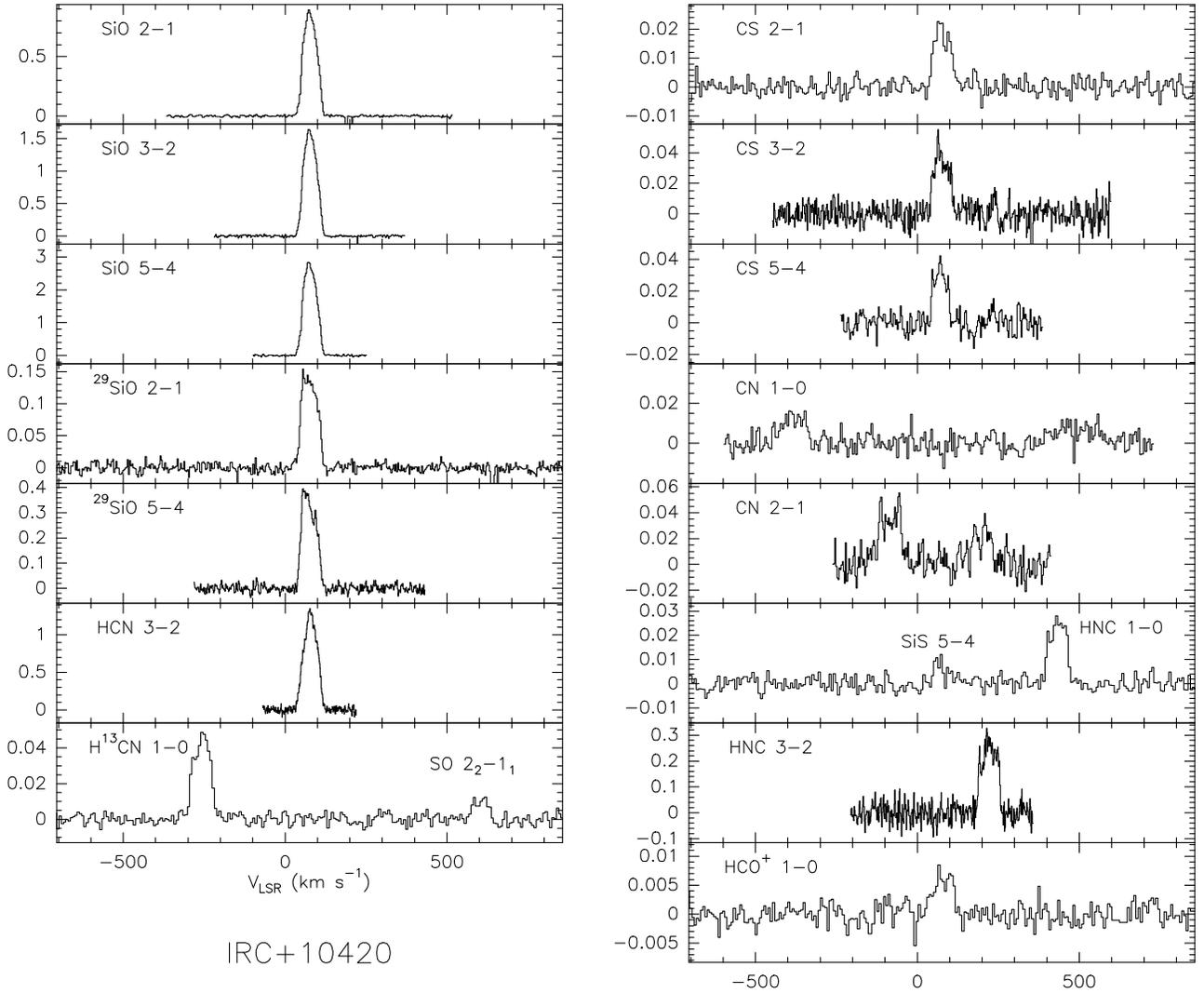}
 \caption{Observed spectra towards IRC\,+10420. The intensity scale is given in units of
 main-beam brightness temperature (K).}
\end{figure*}

In order to compare the line
properties in these sources with those in other evolved circumstellar envelopes,
we have also observed the O-rich AGB stars RX Boo and TX Cam, the C-rich AGB star 
IRC\,+10216, the C-rich PPN CRL\,2688, and the young C-rich PN NGC\,7027. 

The coordinates of the star and some stellar properties of the program stars
are summarized in Table 1. In IRC\,+10216 and CRL\,2688, that are intense emitters, some other lines also appeared within the simultaneous receiver bands of the above lines.

\begin{table*}[t!hpb]
\centering
\caption{Program stars. References: (H) Hipparcos, {(1) Jones et al. (1993)}, 
(2) Bujarrabal et al.\ (1994a), (3) Skinner et al.\ (1998), (4) Jourdain de 
Mouizon et al.\ (1990). }

\begin{tabular} {l c c c c c}
\hline\hline
Name&Observation &coordinates&$V_{\rm LSR}$&$D$&Comments\\
    &$\alpha$(2000)&$\beta$(2000)&(km s$^{-1}$)&(kpc)&\\	
AFGL\,2343&19 13 58.6&00 07 32&98&5.6$^H$& Yellow hypergiant\\
IRC\,+10420&19 26 48.0&11 21 17&76&5$^{1}$& Yellow hypergiant\\
\hline
IRC\,+10216&09 47 57.4&13 16 44&-26&0.2$^{2}$&C-rich AGB star\\
CRL\,2688&21 02 18.8&36 41 38&-35&1.2$^{3}$&C-rich PPN\\
NGC\,7027&21 07 01.6&42 14 10&26&1$^4$&C-rich Young PN\\
RX\,Boo&14 24 11.6&25 42 13&-2&0.2$^{2}$&O-rich AGB star\\
TX\,Cam&05 00 50.4&56 10 53&9&0.35$^{2}$&O-rich AGB star\\
\hline
\end{tabular}

\end{table*}

Our observations were performed in June 2000. 
SIS 3, 2, and 1mm receiver bands were used, often
simultaneously. The receivers were tuned always in SSB
mode. From frequent pointing measurements, we expect errors $\sim$
3''. The spatial resolution
is 12-13'' at 1.3 mm (taking into account the effects of pointing
errors) and 22-26'' at 3 mm. Beam efficiencies ranged between $\sim$
0.75 at 3mm wavelength and $\sim$ 0.5 at 1mm. 

The data presented here are calibrated in units of main beam 
Rayleigh-Jeans-equivalent antenna temperature, $T_{\rm \rm mb}$, using the 
chopper-wheel method by observing hot (ambient) and cold loads 
(liquid nitrogen). The atmospheric conditions were
good, with zenith opacities \lsim\ 0.2 at $\lambda$=1mm. 
In addition, some results were compared with 
previous observations by Bachiller et al.\ (1997a,b), Bujarrabal et
al.\ (1994a), and the catalog of standard line intensities for the 30m
telescope (Mauersberger et al.\ 1989) in order to improve calibration. 
From the comparison between our different spectra and with respect with 
previous data, we expect calibration uncertainties of about 20\%.

The detected spectra in our main sources, IRC\,+10420 and AFGL\,2343, are 
shown in Figs.\ 1 and 2. Note that, in the case of 
CN $N$=2--1, only the two main groups of line components were observed, corresponding to 
$J$=5/2--3/2 and $J$=3/2--1/2. In Figs.\,A.1-A.7 (in electronic version 
only) we present the spectra of the other observed sources: CRL\,2688, NGC\,7027, 
RX\,Boo, TX\,Cam and IRC\,+10216. As baseline profile, only straight lines 
were subtracted.

\section{Observational results}

We have obtained molecular line data of two YHGs, IRC\,+10420 and AFGL\,2343,
as well as for other sources, see Table\ 1. 

In Table 2 we show a summary of
the new observational results: peak intensity (K), rms noise
and profile area (K \kms). These parameters were in most cases calculated for a spectral
resolution of 1 MHz for the 3mm lines and of 2 MHz for the 2mm and 1mm
lines. The actual observations were performed with various
spectrometers, whose resolutions were degraded if necessary to
calculate these parameters. The upper limits are 5-$\sigma$ level,
calculated as follows: 
$Integrated\,Area < 5\,\sigma\,\Delta V /\sqrt{N}$, being $\sigma$ the rms noise 
for the observed spectrum, $\Delta V$ the equivalent velocity width, and 
$N$ the number of channels within
this velocity width. For the 
calculation of $\Delta V$ in the case of CN, we take the equivalent width 
of CN in IRC\,+10216 (Bachiller et al 1997b) as the sum of the areas of all 
the hyperfine transitions divided by its $T_{\rm mb}$. This CN equivalent 
width of IRC\,+10216 is renormalized multiplying it by the ratio between 
$\Delta V$ for $^{13}$CO $J$=2--1 in the star for which the limit is being 
calculated and $\Delta V$ for $^{13}$CO in IRC+10216.

We have also considered the observations of various molecules from
Bujarrabal et al.\ (1994a) and Bujarrabal, Alcolea \& Planesas (1992),
 $^{13}$CO $J$=1--0 and 2--1 data from Bujarrabal et al.\ (2001), CN
 $N$=1--0 and 2--1 data from Bachiller et al.\ (1997b), and HCO$^+$ $J$=1--0
 data from Bachiller et al.\ (1997a), S\'anchez Contreras el al.\ (1997),
 and Lucas \& Gu\'elin (1990). Note the interest of using data obtained
with the same telescope, the IRAM 30m dish, in order to better interpret the line 
ratios. Taking into account
data from the sources in these papers, the total sample of objects which data
are used in our analysis, together with the YHGs, is :

\noindent
{ O-rich AGB stars and red supergiant stars (RSGs)}: RX\,Boo, TX\,Cam, R\,Cnc, RS\,Cnc, VY\,CMa, R\,Cas, $o$\,Cet,
 NML\,Cyg, W\,Hya, R\,Leo, VX\,Sgr, IK\,Tau, RT\,Vir, IRC\,--10529, IRC\,+10011, 
OH\,26.5+0.6, OH\,44.8-2.3.

\noindent 
C-rich AGB stars: LP\,And, UU\,Aur, U\,Cam, Y\,CVn, S\,Cep, V\,Cyg, UX\,Dra, V\,Hya, 
CIT-6, CRL\,865, CRL\,3068, IRC\,--10236, IRC\,+10216, IRC\,+20370, IRC\,+30374, IRC\,+60144.

\noindent
O-rich PPNe: OH\,17.7-2.0, OH\,231.8+4.2, IRAS\,17436+5003.

\noindent
C-rich PPNe: CRL\,618, CRL\,2688, NGC\,7027, IRAS\,07134+1005, IRAS\,19500-1709.

{ Note that the RSGs VY\,CMa, NML\,Cyg, and VX\,Sgr are also
included, but very few data on them are available.
From the point of view of the line intensity ratio comparison,
and also due to the lack of data for these objects, 
both RSGs and O-rich AGBs behave in a similar way.
Also, NGC\,7027, a C-rich young PN, is considered in the following discussion as a PPN.}

\begin{table*}[t]
\centering
\caption{ Peak T$_{\rm mb}$, rms noise, and integrated area from our 
observations. The resolution is 2 MHz for the 1mm and 2mm data, and 
1MHz for the 3mm data. $^*$: 2MHz resolution. $^{**}$: 4MHz resolution.
$\sim$: Tentative detection. $^{13}$CO data from Bujarrabal et al.\ (2001).}
 \scalebox{0.85}{

\begin{tabular} {l c c c c c c c c }

\hline\hline
Line &	 & 		AFGL\,2343 &	 IRC\,+10420 & 	IRC\,+10216 & CRL\,2688 & NGC\,7027 & 
RX Boo & TX Cam  \\
\hline%
$^{13}$CO $J$=1--0 & peak $\pm$ sigma (K)&0.271 $\pm$ 0.007  &0.10 $\pm$ 0.01	  &	  &	  &	  &	 
& \\
   		 & area (K \kms)  &14.15	  &6.12	  &	  &	  &	 & 	 &\\
\hline
$^{13}$CO $J$=2--1 &	&1.0 $\pm$ 0.04 &0.47 $\pm$ 0.024  &4.2 $\pm$ 0.15&2.9 $\pm$ 0.07&&0.15 $\pm$ 0.06&0.22 $\pm$  0.06 	 \\
 		  &	&50.3	  &26.2  &83.3	  &87.7	  &	  &1	  &5.5	 \\
\hline
C$^{18}$O $J$=1--0&	&0.015 $\pm$ 0.004 &	 &	&0.07 $\pm$ 0.01 	& 	&	& 	\\
		&	&0.52	  	&$<$0.29 &	&2.6	  	&$<$0.6	&	& 	\\
\hline
C$^{18}$O $J$=2--1&	&0.08 $\pm$ 0.008	&	&	  &0.40 $\pm$ 0.02 &0.06 $\pm$ 0.02	&	  &     
\\
		  &	&3.8 	 	&$<$0.29&	  &12.4  	&1.3 	        &	  &      \\
\hline
SiO $J$=2--1	&	&0.15 $\pm$ 0.005  &0.9 $\pm$ 0.007  &	  &0.08 $\pm$ 0.02 	&	        &	  &	 
\\
		  &	&6.4	  &44.2  &	  &3.4	  	&$<$0.68  	&	  &	 \\
\hline
SiO $J$=3--2	&	&0.27 $\pm$ 0.007  &1.6 $\pm$ 0.009  &	  &0.26 $\pm$ 0.03  &	  &	  &	 \\
		  &	&11.7	  &80  &	  &8.3	 &$<$0.82 &	  &	 \\
\hline
SiO $J$=5--4	&	&0.59 $\pm$ 0.014  &2.8 $\pm$ 0.02&7.4 $\pm$ 0.3 	&0.35 $\pm$ 0.06&	  &5.22 $\pm$ 0.06 &2.45 $\pm$ 0.03\\
		  &	&17.5	  &139   &183   	&16    &$<$1.26   &56.7   &53    \\
\hline
$^{29}$SiO$J$=2--1&	&0.016 $\pm$ 0.005&0.145 $\pm$ 0.006&0.2 $\pm$ 0.04  &	  &         
&0.238 $\pm$ 0.007&	\\
		  &	&0.95	  	&8.4    &5.9	  &$<$0.8  &$<$0.9	  &3.2	  &	 \\
\hline
$^{29}$SiO$J$=5--4&	&0.064 $\pm$ 0.007&0.38 $\pm$ 0.01&0.5 $\pm$ 0.15**  &    &     &1.2 $\pm$ 0.05 &0.32 $\pm$ 0.024	 \\
		  &	&1.5	  	&20         &12      &$<$1&$<$1.1&12.8      &6.4	 \\
\hline
SO $J$=2$_2$--1$_1$&	&	  &0.012 $\pm$ 0.002*&	  &	  &         &	  &0.023 $\pm$ 0.007	 
\\
		  &	&	  &0.63  	&	  &$<$1.1 &$<$1.1   &	  &0.52	 \\
\hline
HCN $J$=1--0	  &	&0.036 $\pm$ 0.004&	  &	  &6 $\pm$ 0.014	  &	  &	  &	 \\
		  &	&1.9	  &	  &	  &170    &	  &	  &	 \\
\hline
HCN $J$=3--2	&	&0.13 $\pm$ 0.01  &1.3 $\pm$ 0.02&	  &13 $\pm$ 0.06 &1.9 $\pm$ 0.045  &	  &	 \\
		  &	&3.3	  &65.4  &	  &400   &46  &	  &	 \\
\hline
H$^{13}$CN $J$=1--0&	&  	& 0.047 $\pm$ 0.004&	  &2.18 $\pm$ 0.03  &      &  &0.037 $\pm$ 0.006	 \\
		&	&  	& 3.04     	 &	  &63.7  &$<$1.1	&	  &1.1	 \\
\hline
HC$_3$N $J$=10--9&	&	  &	  &	  & 1.36 $\pm$ 0.01 & 		&	  &	 \\
		&	&$<$0.23  &$<$0.26 &	  & 42   &$<$0.62	&	  &	 \\ 
\hline
CS $J$=2--1	&	&0.023 $\pm$ 0.004&0.023 $\pm$ 0.004&7 $\pm$ 0.04&0.97 $\pm$ 0.02 &	  &	     
&	 \\
		&	&0.72	 	&1.2      	&174.3		&29    &$<$0.73	  &$<$0.38	  &	 \\
\hline
CS $J$=3--2	&	&0.04 $\pm$ 0.004	&0.045 $\pm$ 0.004&	  &1.93 $\pm$ 0.02&	  	&	  &	 \\
		&	&1.54  &2.2    &	  &57.3  &$<$0.9	  &	  &	 \\
\hline
CS $J$=5--4	&	& 0.034 $\pm$ 0.006	&0.042 $\pm$ 0.007&	  &2.03 $\pm$ 0.015&   &	  &	 \\
		&	& 1.16 	   	&1.64  &	  &60.0  &$<$1.6 &	  &	 \\ 
\hline
SiS $J$=5--4	&	&	  &0.011 $\pm$ 0.003*	&	  &0.2 $\pm$ 0.01 &	      &	  &	 \\
	        &	&$<$0.23  &0.37      	&	  &5.3	      &$<$0.62&	  &	 \\
\hline
SiS $J$=15--14	&	&	  &	  &	  &0.45 $\pm$ 0.08&	 &	  &	 \\
		&	&$<$1     &$<$1.4  &	  &15    &$<$2.55&	  &	 \\
\hline
CN $J$=1--0	&	& 	 &0.016 $\pm$ 0.004* &	  &2.1 $\pm$ 0.036 &0.46 $\pm$ 0.02&	  &	 \\
		&	&$<$0.31 &2.6        &	  &153.4 &32.50  &$<$0.29	  &	 \\
\hline
CN $J$=2--1	&	&$\sim$0.012 $\pm$ 0.004** &0.05 $\pm$ 0.009&	  &3.9 $\pm$ 0.2 &1.3 $\pm$ 0.03& 
&	 \\
		&	&$\sim$0.67	       &4.6	   	&	  &192.7     &88.2      &$<$0.21 &\\
\hline
HNC $J$=1--0	&	&0.008 $\pm$ 0.003 &0.026 $\pm$ 0.004        &	  &0.77 $\pm$ 0.01& $\sim$0.03 $\pm$ 0.01*&  &	 \\
		  &	&0.31	 &1.7  		 &	  &22.4  &$\sim$0.93 &	  &	 \\
\hline
HNC $J$=3--2s	&	&            &0.28 $\pm$ 0.03&	  &3.4 $\pm$ 0.08 & 	 &	  &	 \\
		  &	&$<$1	     &15.6  &	  &109.4  &$<$2.7 	&	  &	 \\
\hline
HCO$^+$ $J$=1--0 &	&0.008 $\pm$ 0.002 &0.007 $\pm$ 0.002*	&&0.03 $\pm$ 0.005*&1.34 $\pm$ 0.01 
&	  &	 \\
		  &	&0.4	  	 &0.48  	&	  &0.86  	&29.4  &$<$0.12  &$<$0.23	 \\
\hline
HCO$^+$ $J$=3--2 &	&	  &	  &	  &0.15 $\pm$ 0.045&3.2 $\pm$ 0.09  &	  &	 \\
		  &	&$<$1     &$<$0.9 &	  &5.9	  &92  &	  &	 \\		

\hline

\end{tabular}

 }

\end{table*}
\begin{figure*}
\centering
 \includegraphics[angle=0, width=17cm ]{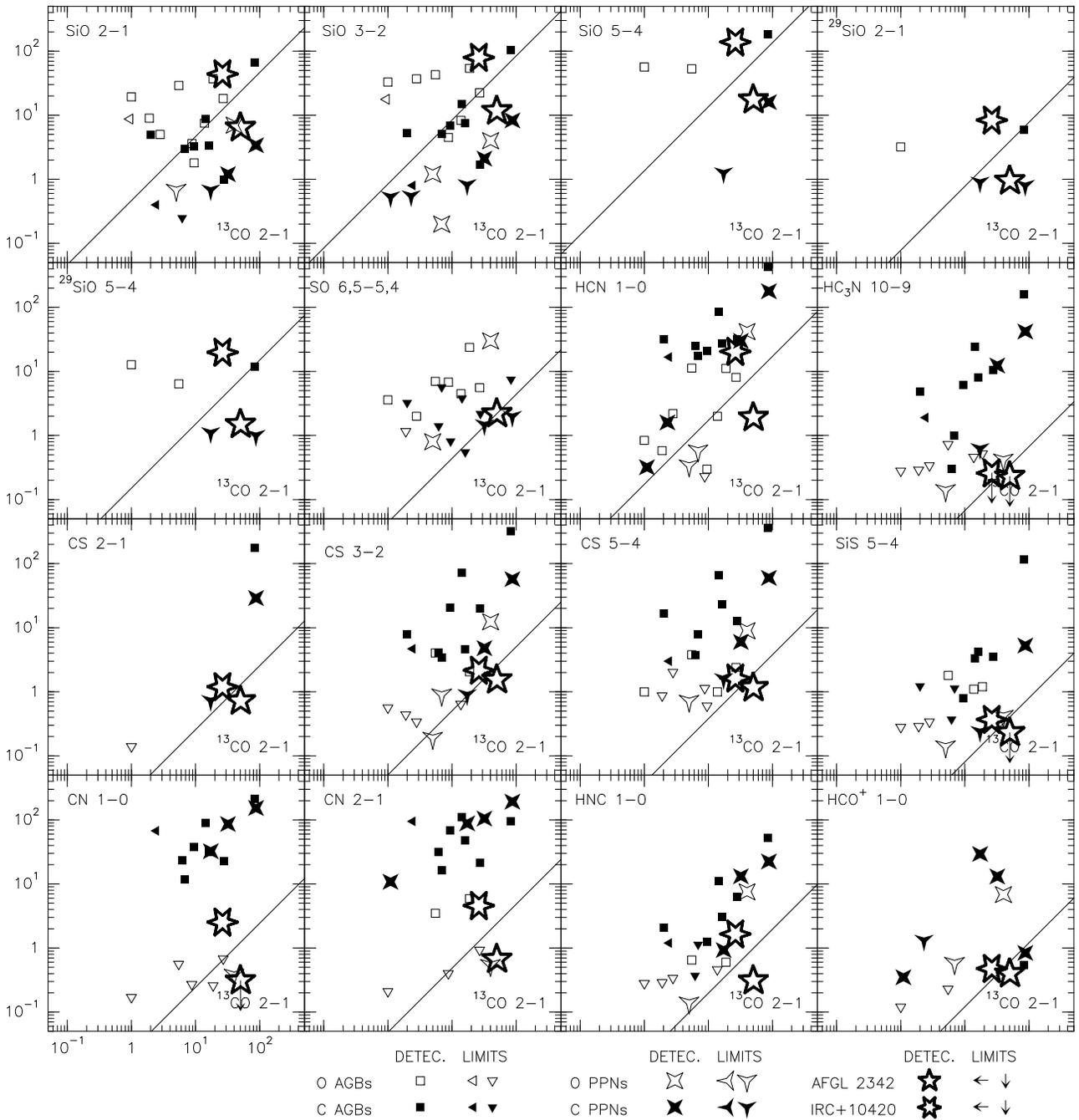}
 \caption{Profile area diagrams showing some representative line transitions 
versus $^{13}$CO $J$=2-1.}
\end{figure*}

\begin{figure*}
\centering
 \includegraphics[angle=0, width=17cm ]{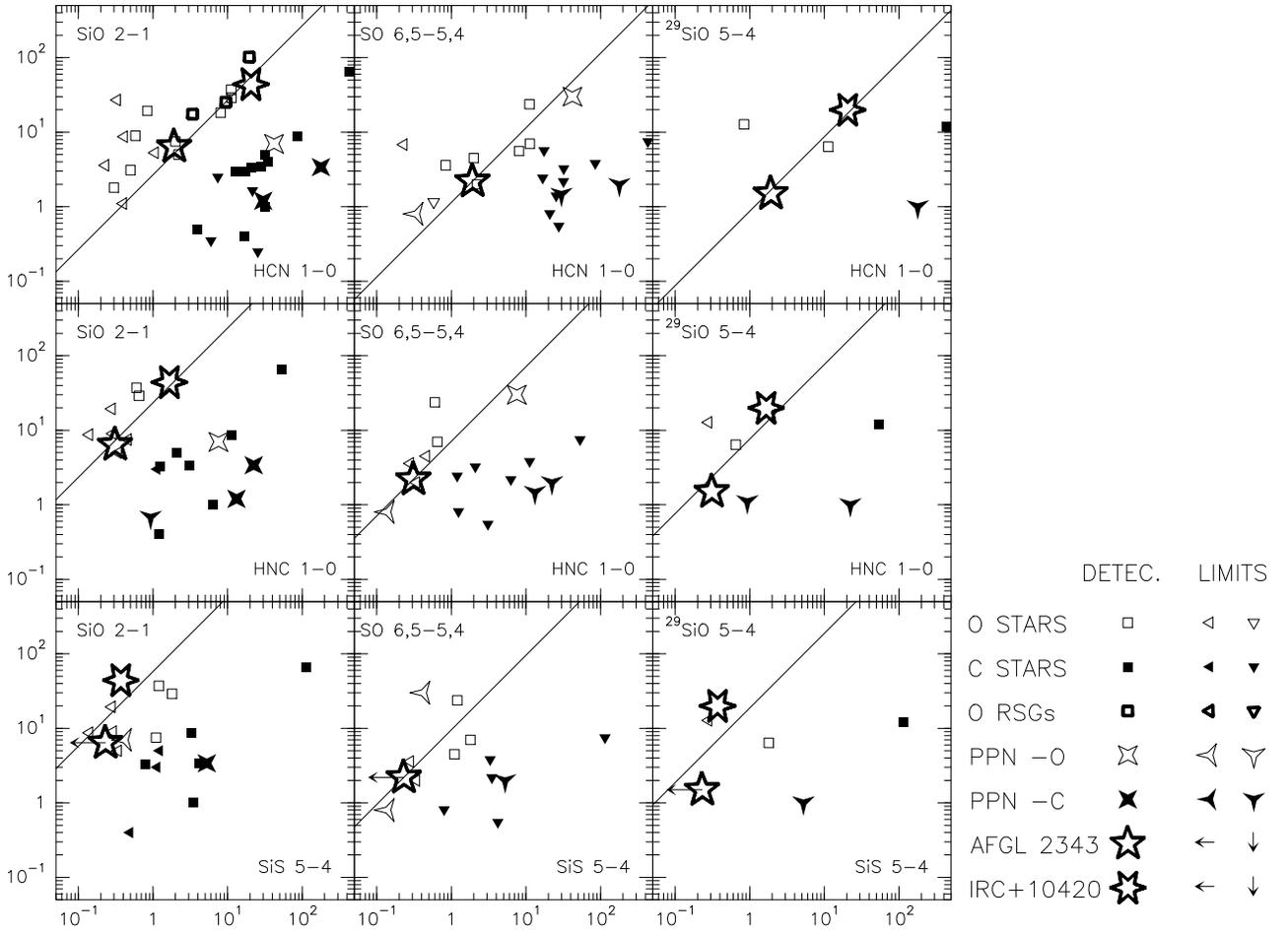}
\caption{Profile area diagrams for some pairs of transitions of O-rich 
molecules versus some of C-rich ones.}
\end{figure*}

\begin{figure*}
\centering
 \includegraphics[angle=0, width=17cm ]{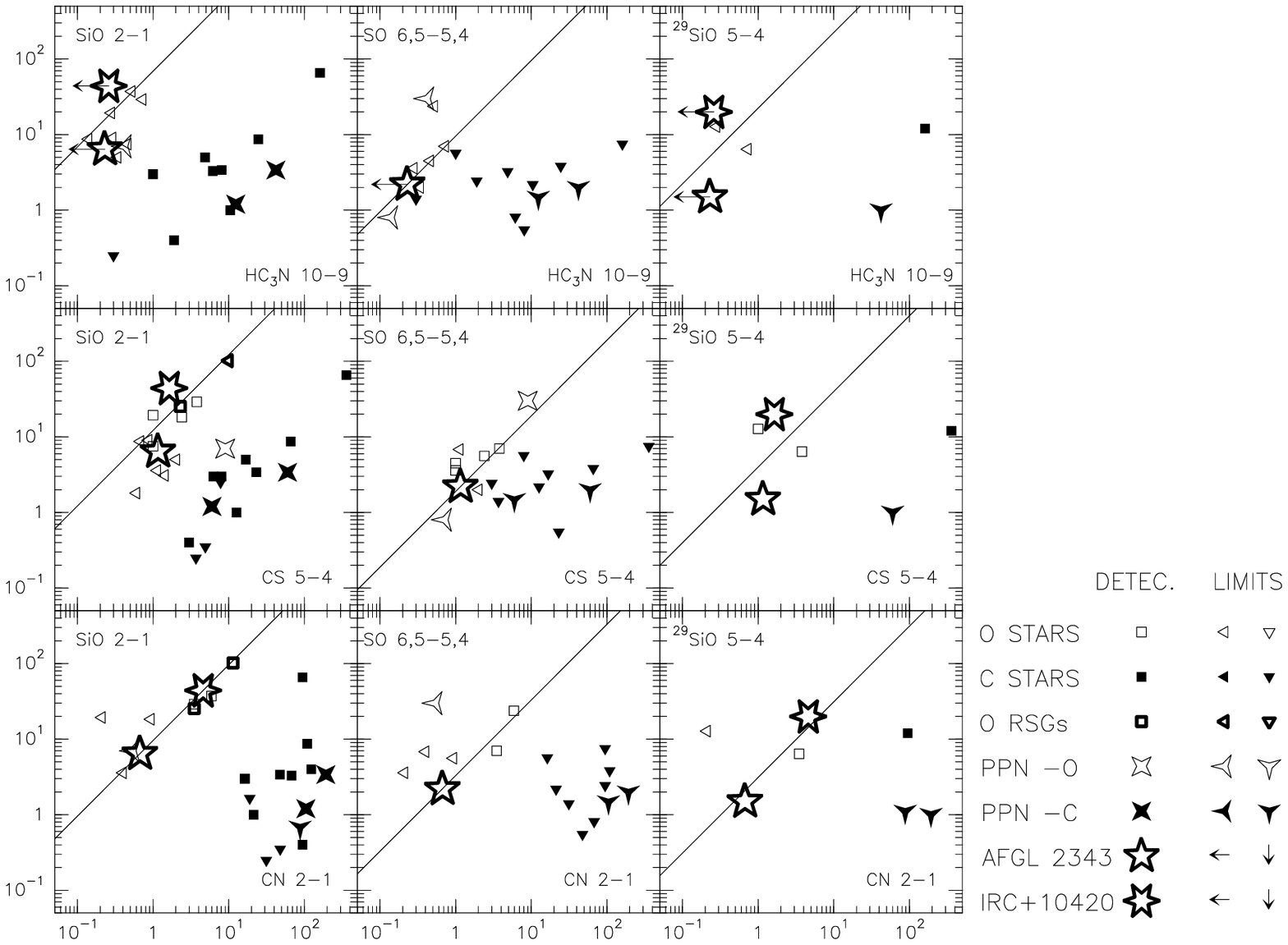}
 \caption{Same as Fig. 4 for other pairs.}
\end{figure*}

\subsection{Line intensity comparison}

In order to compare the molecular emission in YHGs with that of other
CSEs, we have represented in Figs.\ 3 to 5 
integrated intensities of several pairs of relevant lines. Open symbols 
represent O-rich objects, and filled symbols C-rich objects. Squares and
 triangles respectively represent detections and limits for the AGB stars (AGBs).
{ The RSGs are plotted as O-rich AGBs with thicker symbols.} 
Four-peak asterisks and three-peak asterisks represent, respectively, 
detections and limits for PPNe. These polygons with three vertices 
 have one of them  pointing leftwards or downwards, 
indicating for which transition we have a limit. Objects 
with limits for the two represented lines are not plotted.
For the YHGs, a five-peak open asterisk represents AFGL\,2343, a six-peak 
asterisk is used for IRC\,+10420, and the limits are represented by an 
arrow from the center of the symbol. 
To make the comparison easier with the other objects,
straight lines (hereafter average lines) represent the points with a ratio equal 
to the geometric average of the profile area ratios for the two YHGs.

In Fig.\,3 we show ratios between integrated line intensities for \trece\ and other
abundant molecules. In Fig.\,4 and 5 we show line ratios between pairs
of molecules other than \trece.
In Fig. 3 we can see that, in general, AFGL\,2343 and IRC\,+10420 lay 
in significantly different regions of the diagrams. For IRC\,+10420
these line ratios are comparable to these typical in O-rich AGB stars.
For AFGL\,2343, however, they are weaker by factors ranging between $\sim$ 2 and
10. On the other hand, 
in Figs. 4 and 5 both YHGs fall very close to the average line and to the region 
of the diagrams occupied by O-rich AGB stars, and their line ratios are clearly different
from those of C-rich stars and PPNe. 
The situation of AFGL\,2343 in Fig.\,3 could be due to an overabundance 
in  $^{13}$CO, an underabundance of the rest of the molecules, or to a 
small mass of the emitting regions for these 
molecules. Note that, 
for optically thin lines, the ratio of the profile areas
depends on the abundance of the molecule and the mass of 
the emitting region. We can conclude from these diagrams that the YHGs 
show, as expected, O-rich chemistry similar to that of the O-rich AGB stars. 

The relative weakness of most molecular lines with respect to $^{13}$CO
in AFGL\,2343 is particularly noticeable for $^{28}$SiO, $^{29}$SiO, HCN, 
CN and HNC. The $^{28}$SiO/$^{13}$CO intensity ratio in AFGL\,2343 is 
similar to that of PPNe, and lower than those ratios obtained for AGBs, O-rich or 
C-rich. We also confirm the previously-reported low SiO intensity 
in PPNe.

The HCN intensity of AFGL\,2343, relative to $^{13}$CO, lays under 
all O-rich AGBs. The HNC/$^{13}$CO and CN/$^{13}$CO line intensity 
ratios in this source are the lowest among all sources in our sample 
in which HCN or CN is detected.

To go a step further in the comparison between the molecular line intensity 
in these kinds of stars, we will now use profile areas corrected for 
distance. In general, for the same intrinsic luminosity, the 
observed brightness intensity of sources at different distances from the observer 
follows a $\sim D^{-2}$ law, where $D$ is the distance. Multiplying the 
intensities obtained by $D^2$, in kpc, we set, in some way, all the stars 
at the same distance of 1\,kpc. The distances for the YHGs were taken 
from de Jager (1998) in the case of IRC\,+10420, and from Hipparcos 
parallax measurements for AFGL\,2343, which, despite of its low accuracy, 
is compatible with the luminosity of a YHG.
For the rest of stars the distances are give in Table 1 or were taken from Bujarrabal 
et al.\ (1994a).

The result of the distance correction is presented in Fig.\ 6, for the following 
representative transitions for each chemistry type: SiO $J$=2--1 (O-rich) and 
$J$=3-2, HCN $J$=1--0 and HNC $J$=1--0 (C-rich), all versus $^{13}$CO.
We found, in order of increasing 
line intensities, first the AGBs, then PPNe, and finally the YHGs.
This difference is higher in the case of SiO, an O-bearing molecule, than 
for HCN and HNC, which are C-bearing. The reason for this behavior is 
that, as we have seen, the YHGs show O-rich chemistry.

We note that, although the use of distance corrected areas shows a 
difference between YHGs and the other objects, it depends directly on 
the quality of the distance determination. If, as claimed by 
Josselin \& L\`ebre (2001), the distance for AFGL\,2343 is smaller than 
that used here, it would lay in the AGB or PPN region of the diagrams in Fig.\ 6.

\subsection{Line widths}

The equivalent velocity width, $\Delta V$, provides an independent comparison 
between YHGs, AGBs and PPNe.
We have used as reference line $^{13}$CO $J$=1-0, for which we
 measured $\Delta V$ for all objects.

The mean values found are: 

\noindent
AGBs:  $\Delta V$\,=\,(22 $\pm$ 9) km s$^{-1}$

\noindent
PPNe: $\Delta V$\,=\,(29 $\pm$ 9) km s$^{-1}$

\noindent
YHGs:  $\Delta V$\,=\,(53 $\pm$ 3) km s$^{-1}$

In Fig.\ 7 we show a histogram of the obtained equivalent line width.
This statistical approach to $\Delta V$ in these evolved objects signalizes
a major difference between YHGs and  AGBs and PPNe. Note that 
this comparison is distance independent. This 
result gives us another reason to claim that AFGL\,2343 is probably a YHG.

{ Due to the lack of data for the RSGs we can not obtain an estimate of the equivalent width using this method. From $^{12}$CO $J$=2--1 line data 
(Cernicharo et al. 1997) a main value of 55\,\kms\ is found for RSGs, similar to that of YHGs.
This is a confirmation of the evolutionary connection between these objects
(see e.g. de Jager 1998).}

Note that, for many PPNe, the lines show a core (region with low velocity) 
and wings (high velocity). The emission in the profile wings comes from regions with a high 
expansion velocity, even larger than 100 km\,s$^{-1}$ (see e.g. Bujarrabal
et al. 2001), but weaker than that of the profile core.
The equivalent width is therefore dominated by the core region and it is not
very large. In any case, the shape of the line profiles in PPNe is very different
from that of AFGL\,2343 and IRC+\,10420, which do not show wings (see 
Figs.\ 1 \& 2).


\section{Abundances}

\subsection{Abundance estimate method}

The formulation used to calculate the abundances from the measured intensities 
is similar to that used by Bujarrabal et al.\ (2001). We suppose that 
the populations of all the rotational levels can be described by a single 
temperature $T_{\rm rot}$. We do not assume optically thin emission. 
The abundance of a molecule can be estimated from the equation:

\begin{eqnarray}
X =  ln \left[\frac{1}{1-\frac{T_{\rm mb}}{S(T_{\rm rot})} \frac{\Omega_B}{\Omega_S}}\right]
\frac{8\pi\nu_o^3\, Q}{c^3 g_u A}\, \frac{e^{\,E_l/kT_{\rm rot}}}{(1-e^{-h\nu_o/kT_{\rm rot}})}\,\times 
\nonumber \\ \times \frac{\Omega_S D^2 \,m(H_2)}{M} \Delta V   ,
\end{eqnarray}
\noindent
where $X$ is the fractional abundance, $Q$ is the partition function, 
$\Omega _B$ is the HPBW beam size, $\Omega_S$ is the size of the 
source, $\Delta V$ is the equivalent line width of the emission (in cm 
s$^{-1}$), $g_u$ is the degeneration of the upper state, $A$ is the 
Einstein coefficient, $E_l$ is the energy of the lower state of the 
transition, and $M$ is the mass of the emitting gas. For linear 
molecules, $E_l =h B_o J_l(J_l+1)$, where $B_o$ is the rotational 
constant. Note that, in general, the source is much smaller than the 
beam and the lines are optically thin, therefore the abundance does not 
depend on $\Omega_S$.

In Eq.\ 1 the logarithm term appears to account for optically thick 
emission. However this approximation is only useful for moderate 
opacities, $\tau$. For values of the optical depth larger than $\sim$ 2
, the results become too strongly affected by observational and model 
uncertainties. In other words, for high $\tau$, the emission will only 
come from the external layers of the envelope, since we are unable to 
detect the emission from the inner zones, and the derived abundance 
will just be a lower limit. So, when our calculations indicate an optical 
depth larger than 2, we just give the value corresponding to $\tau = 2$ as 
the lower limit to the abundance (this happens for all the transitions
of SiO, $^{29}$SiO, and HCN in IRC\,+10420) . We also note that this expression 
becomes merely approximate for $\tau \gsim 1$ , since we use mean values 
of $\tau$ within the profile to avoid the integral in the observed velocity. 

Due to the low values of $T_{\rm rot}$ we found (see below), we 
decided to numerically calculate the partition function with high 
accuracy, rather than to use approximations. The sum was pursued 
until the next term had reached a very low tolerance level. In the 
case of SO, $Q$ was calculated using the energy levels given by Omont 
et al.\ (1993).

CN has fine and hyperfine structure (see Figs. 1 \& 2). In order 
to simplify the calculation of $Q$, we reduce this case to a two $N$-level system
by recombination of their Einstein coefficients.

\begin{eqnarray}
A_{\rm u\,l} = \frac{1}{\sum_{\rm u} g_u} \sum_{\rm \alpha', \alpha} g_u\,A_{\rm u\,\alpha', l\,\alpha}  .
\end{eqnarray}

This procedure can be safely used only for optically thin emission, where 
the relative intensity of each line is proportional to the Einstein coefficient, $A$.

The radius of the molecule-rich nebulae around AFGL\,2343 and IRC\,+10420 
are taken from results from interferometric CO maps by Castro-Carrizo et al.\ (2007). 
We will assume that the $^{13}$CO emission extends as far as  
$^{12}$CO, while the emission from the rest of the  molecules is 
supposed to come from the densest part of the shell. In fact, in 
the case of IRC\,+10420, the SiO emission is known to appear in 
this dense region (Castro-Carrizo et al.\ 2001). The mass of the 
emitting region is also derived from the results obtained by 
Castro-Carrizo et al.\ (2007). These data are summarized in table 3.

\begin{figure}
\centering
 \resizebox{\hsize}{!}{\includegraphics{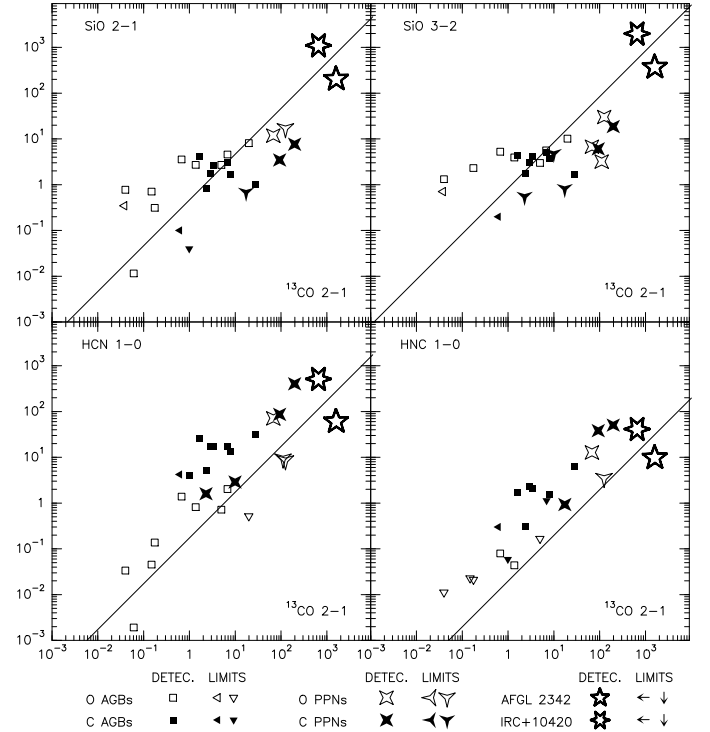}}
 \caption{Distance-corrected profile areas diagram for some 
representative lines versus $^{13}$CO $J$=2-1}
\end{figure}

\begin{table}[t!hpb]
\centering
\caption{Assumed radii and mass of the emitting regions, based on the 
$^{12}$CO data from Castro-Carrizo et al.\ (2007)}
\scalebox{0.8}{

\begin{tabular} {l | c c c | c c c}
\hline\hline
Molecule	&$R_{ \rm in}$(cm)&$R_{ \rm out}$(cm)	&$M$ (\ms)&$R_{ \rm in}$(cm)&$R_{ \rm out}$(cm)	&$M$ (\ms)\\
\hline
&&AFGL\,2343 &&&IRC\,+10420\\
\hline
$^{13}$CO	&1\,10$^{15}$&5\,10$^{17}$		&4.02&3\,10$^{16}$&5.2\,10$^{17}$		&0.7\\
Others		&1\,10$^{15}$&2.5\,10$^{17}$		&3.31&3\,10$^{16}$&1.24\,10$^{17}$	&0.24\\
\hline
\end{tabular}

}

\end{table}

\begin{figure}
\centering
 \resizebox{7cm}{!}{\includegraphics{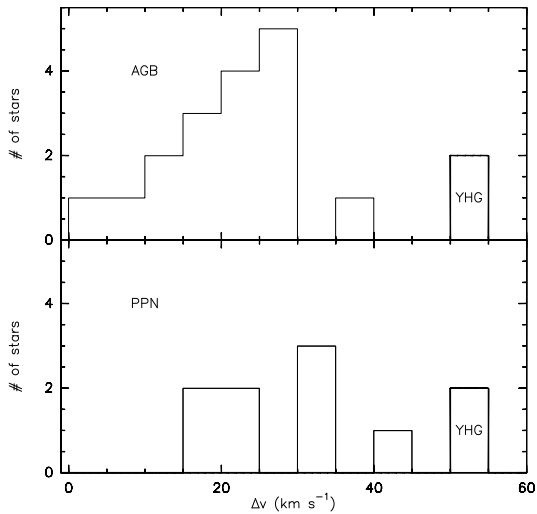}}
 \caption{Histogram of the profile equivalent line width. 
Upper panel shows the comparison between 
AGBs and YHGs. Lower panel refers to PPNe versus YHGs}
\end{figure}

We estimate $T_{\rm rot}$ from the line intensity ratio for the molecules 
in which we have observed more than one transition. 
Sometimes, the value of $T_{\rm rot}$ found in this way was slightly 
lower than the observed brightness temperature, after correcting the 
observed $T_{\rm mb}$ for the dilution factor, $\Omega_S/\Omega_B$. 
In these cases, we imposed $T_{\rm rot}$ to be equal to the 
brightness temperature.

For the molecules with only one observed transition or with only one 
detection, we took an average rotational temperature, which was 
found to be $\sim$ 10 K for 
AFGL\,2343 and $\sim$ 11 K for IRC\,+10420. In some cases,  where we 
have a limit for the transition with the highest $J$, we obtained that the upper limit 
to the abundance from the non-detection is lower than the abundance 
derived from the detected line, using the mean $T_{\rm rot}$. In this 
case, we decrease the rotational temperature
until both values of the abundances derived are equal, being an upper 
limit for $T_{\rm rot}$. 

\begin{table}[t!hpb]
\renewcommand{\footnoterule}{}
\centering
\caption{Mean abundances and rotational temperatures derived for 
AFGL\,2343 and IRC\,+10420. $^1$: forced to be equal to the average 
$T_{\rm rot}$ value.  $^2$: $T_{\rm rot}$ imposed to be equal to brightness temperature.
$^3$: from intensity upper limit; see text for details.}
\scalebox{0.9}{

\begin{tabular} {l | c c | c c}
\hline\hline
Line&	$\langle\,X\,\rangle$(AFGL\,2343)&	$T_{\rm rot}$(K)&	$\langle\,X\,\rangle$(IRC\,+10420)&	$T_{\rm rot}$(K)	\\
\hline
&&&&\\	
$^{13}$CO	& 2.6\,10$^{-5}$\,&9	&4.7\,10$^{-5}$&10	\\
C$^{18}$O	&1.5\,10$^{-6}$	&20	&$<$6.6\,10$^{-6}$&11$^1$\\
CN		&3.6\,10$^{-9}$&6	&3.8\,10$^{-7}$&6	\\
CS		&8.3\,10$^{-9}$	&10	&1.9\,10$^{-7}$&9	\\
H$^{13}$CN	&		&	&2.9\,10$^{-7}$&11$^1$	\\
HC$_3$N		&5.5\,10$^{-9}$	&10$^1$	&5.7\,10$^{-8}$&11$^1$	\\
HCN		&1.1\,10$^{-8}$	&7	&$>$2.2\,10$^{-6}$&36$^2$	\\
HNC		&2.0\,10$^{-9}$&$\le$\,8$^3$	&1.6\,10$^{-7}$&18	\\
HCO$^+$		&2.0\,10$^{-9}$&$\le$\,7$^3$	&2.7\,10$^{-8}$&$\le$\,6$^3$	\\
SiO		&5.4\,10$^{-8}$	&11	&$>$1.3\,10$^{-5}$&79$^2$	\\
SiS		&8.5\,10$^{-9}$ &10$^1$	&1.5\,10$^{-7}$&11$^1$	\\
SO		&		&	&1.6\,10$^{-6}$&11	\\ 
$^{29}$SiO	&6.9\,10$^{-9}$	&8	&$>$1.0\,10$^{-6}$&15$^2$	\\
&&&&\\
\hline
\end{tabular}

}

\end{table}

\subsection{Abundance results}

We present the results of our abundance calculations in Table 4. We 
also show the derived rotational temperatures. The temperatures are 
systematically higher for the molecules that show high optical depths, 
as expected when the excitation temperatures are lower than the kinetic 
temperature. 

We deduce the mean densities for the CO emitting region
from the data in Table 3:  
$n \sim 5 \times 10^{3}$ cm$^{-3}$ for AFGL\,2343, and $n \sim 8 
\times 10^{2}$ cm$^{-3}$ for IRC\,+10420. The mean densities derived for the emitting
regions of the other molecules are $n \sim 4 \times 10^{4}$ cm$^{-3}$ for AFGL\,2343 and
$n \sim 2 \times 10^{4}$ cm$^{-3}$ for IRC\,+10420.
These relatively low densities are compatible with the idea 
that most lines, except in particular those of CO, should be underexcited. 

AFGL\,2343 shows a general underabundance in all the molecules but $^{13}$CO.
The abundances of molecules other than $^{13}$CO 
are in general more than ten times higher in IRC\,+10420 than in AFGL\,2343. 
AFGL\,2343 shows an abundance of $^{13}$CO comparable to that typical of AGB stars, 
while this abundance in IRC\,+10420 is higher by a factor 2.

The emission of molecules in AFGL\,2343, apart from $^{13}$CO, could come 
from a region smaller than that assumed here (Sect.\,4.1), containing significantly
less mass.
This would result in an increase in the derived molecular abundances (in that region) 
for AFGL\,2343. In fact 
the emission of such species in AGB envelopes usually comes from regions at a distance 
from the star of several 10$^{16}$\,cm, although the SiO emission from IRC\,+10420
is confirmed to appear at a distance of about 10$^{17}$\,cm. 
We note the difference in the assumed radii of the dense region for both 
stars (see Table 3), being that of AFGL\,2343 twice that of IRC\,+10420. 

\subsection{Comparison with previously published results.}

With the results in Table 4, we are able to compare the abundances here 
calculated with other published results for YHGs, like those obtained in 
Bujarrabal et al.\ (1994a) and Bachiller et al.\ (1997b). 
In these papers the distance assumed for IRC\,+10420 is 3.4 kpc and for 
AFGL\,2343 is 6 kpc. To compare the results of the abundance calculations 
first we must correct for the effects of the different assumed distances. 

The average difference is $\sim$20 for IRC\,+10420 and around $\sim$300 for AFGL\,2343.
In order to find out the origin of these discrepancies we must focus on the assumptions 
made in the different calculations. Both Bujarrabal et al.\ (1994a) and Bachiller 
et al.\ (1997b) used the same method, assuming optically thin emission. In fact, 
Eq.\ 1, here used to calculate abundances, would become that used by Bujarrabal et 
al.\ (1994a,b) and Bachiller et al.\ (1997b) if we assume, as in those papers, 
optically thin emission and a negligible background temperature ($T_{\rm bg}\sim 0$). 
For those cases with optically thin emission, the differences 
in the calculations can only be due to the assumed values of the emitting mass and 
of $T_{\rm rot}$. For optically thick cases, the differences also depend on the source size and opacity.

In Table 5 we can see the radii 
used and the masses obtained for the emitting regions by Bujarrabal et 
al. (1994a) and Bachiller et al. (1997b). The radii assumed by these authors
for regions emitting in molecules other 
than CO  in YHGs are significantly lower than the values obtained from PdB maps (Sec. 3), leading to 
smaller masses and, therefore, to higher abundances. Indeed, the 
discrepancies found in the abundances are almost fully explained by these differences 
in the masses adopted for the emitting region.

\begin{table}[t!hpb]
\centering
\caption{Radii and mass of the emitting gas for each molecule used  in Bujarrabal 
et al.\ (1994a) and in Bachiller et al. (1997b). *:\ Shell, $R_{\rm in}=3.1\,10^{16}$cm.}
\scalebox{1}{

\begin{tabular} {l | l l | l l}
\hline\hline
Molecule	&$R_{\rm out}$(cm)	&$M$ (\ms)&$R_{\rm out}$(cm)	&$M$ (\ms)\\
\hline
&AFGL\,2343 &&IRC+10420\\
\hline
$^{13}$CO	&10$^{17}$		&0.2&10$^{17}$		&0.06\\
SiO		&2\,10$^{15}$		&4.2\,10$^{-3}$&2\,10$^{15}$		&1.2\,10$^{-3}$\\
CN		&&&1.1\,10$^{17}$*	&0.05\\
Other		&10$^{16}$		&0.02&10$^{16}$		&6\,10$^{-3}$\\
\hline
\end{tabular}

}

\end{table}

Another important factor in the abundance estimate is the rotational 
temperature. In Bujarrabal et al.\ (1994a,b), a $T_{\rm rot}$ of 20\,K 
is assumed. The temperatures we found here are in most of the cases 
lower than 20\,K. Bachiller et al.\ (1997b), however, estimate the 
rotational temperature from the ratio between CN $N$=2--1 and 
$N$=1--0 emission, and it is similar to the one obtained here for this 
molecular transition. 

In some cases, the ratio between the 
abundances calculated here and those from the cited authors
 is lower than the ratio of the masses. This is explained by the 
difference in $T_{\rm rot}$. Lower rotational temperatures lead to 
higher abundances in the case of transitions with a high $J$, like in 
the case of CS $J$ =3--2 and $J$ = 5--4, which were the transitions 
used to calculate the abundance in Bujarrabal et al.\ (1994a).

\subsection{Comparison with abundances in AGB stars.}

The relative abundances, $X$, determined here (see Table 4) must be compared with the 
standard abundances found in circumstellar envelopes (e.g. Bujarrabal et al.\
1994a), in order to understand the differences between our YHGs and those.

The values of $X$ found for IRC\,+10420 are in full agreement with those 
supposed for the O-rich CSEs, except for the high abundances obtained for
 $^{13}$CO and HCN. In the case of $^{13}$CO, this overabundance is a 
factor 2. Note that the value of $X_{\rm HCN}$ is affected by the 
opacity, as said in Sect.\,4.1, therefore low abundances were
found by Bujarrabal et al.\ (1994), who assumed low optical depths.

In AFGL\,2343, except for $^{13}$CO, the abundances are much lower
 (around 70 times on average) than the standard values for the CSEs around AGBs.
As we mentioned in Sect.\,4.2., such low abundances found in AFGL\,2343 
could be related to the extent
of the circumstellar layer emitting in these molecular lines. 
In AFGL\,2343, the region we assumed to be
molecule rich is as extended as 2.5 10$^{17}$ cm, but there is no
observational confirmation of such a large extent. If the region at
which molecules other than CO are abundant in this source is
comparable to that of IRC\,+10420, then the emitting mass would be
much smaller than the mass assumed in our abundance calculation
($\sim$0.05\ms, see mass estimates by Castro-Carrizo et al. 2007).
This would yield molecular abundances in this region comparable to those usually
found in molecule rich CSEs.


\section{Conclusions}

We observed several molecular transitions of a wide variety of species (see
Sect. 2) in the yellow hypergiants (YHGs) AFGL\,2343 and IRC\,+10420 as well as in 
several AGB stars and PPNe.
Previously published observations were added to 
our sample, to obtain a better comparison between the properties of the different
objects.

From the collected spectra, we have compared the ratios of the integrated area
for several relevant pairs of lines in the YHGs with those in AGBs and PPNe.
We find that the YHGs studied here show, as expected, O-rich chemistry. 
However, the emission of molecules other than $^{13}$CO
in AFGL\,2343 is significantly weaker than that found for the O-rich AGBs and for IRC\,+10420.

The comparison of the ratios of the integrated profile areas corrected by distance has shown
a clear difference between YHGs, whose emission is the more intense, and PPNe and
AGB stars. Another difference, distance independent, appears when comparing equivalent
line widths, which are much higher for YHGs than for the other evolved objects. 
At this respect, we note the absence of wings 
in the line profiles of AFGL\,2343 and IRC\,+10420.

All this supports the idea that AFGL\,2343 is a YHG rather than a PPN, as also deduced from CO maps
(Castro-Carrizo et al. 2007).

We estimated the abundances of the different molecular species 
within the circumstellar envelopes of AFGL\,2343 and IRC\,+10420 using the method
described in Sect.\,4.1. We assumed that molecular emission, 
apart from CO and its isotopes, comes from the innermost dense region found by 
Castro-Carrizo et al.\ (2007) and described in Table 3, which extends to several
 10$^{17}$\,cm. Note that this extent is larger than that usually found for the 
same molecules in AGB stars. Dust shielding would prevent photodissociation of 
these molecules in the further regions assumed for the YHGs.
On the other hand, \trece\ line emission in AGB CSEs and PPNe usually extends as much 
as \doce\ emission; we recall that the \doce\ envelope in our sources has been mapped
by Castro-Carrizo et al. (2007).

The abundance found for $^{13}$CO in AFGL\,2343 is comparable with that usual for
 AGBs, while that of IRC\,+10420 is larger by a factor of 2. 
The abundances of other molecules found in IRC\,+10420 are comparable, 
in general, to those of O-rich AGB stars, but those of AFGL\,2343 are found to be 
around 40 times lower in average. These low abundances could 
be due to the fact that these molecules are in fact abundant only in an inner region
with a relatively low mass.

\acknowledgements
{This work has been supported by the Spanish Ministerio de Ciencia y
Tecnologia and European FEDER funds, under grants AYA2003-7584 and 
ESP2003-04957. The contribution of AC-C was supported by the 6th European 
Community Framework Programme through a Marie-Curie Intra-European Fellowship. }

{}

\appendix

\section{Spectra of other program stars.}

\newpage
\begin{figure*}[h]
\centering
 \resizebox{\hsize}{!}{\includegraphics[angle=-90]{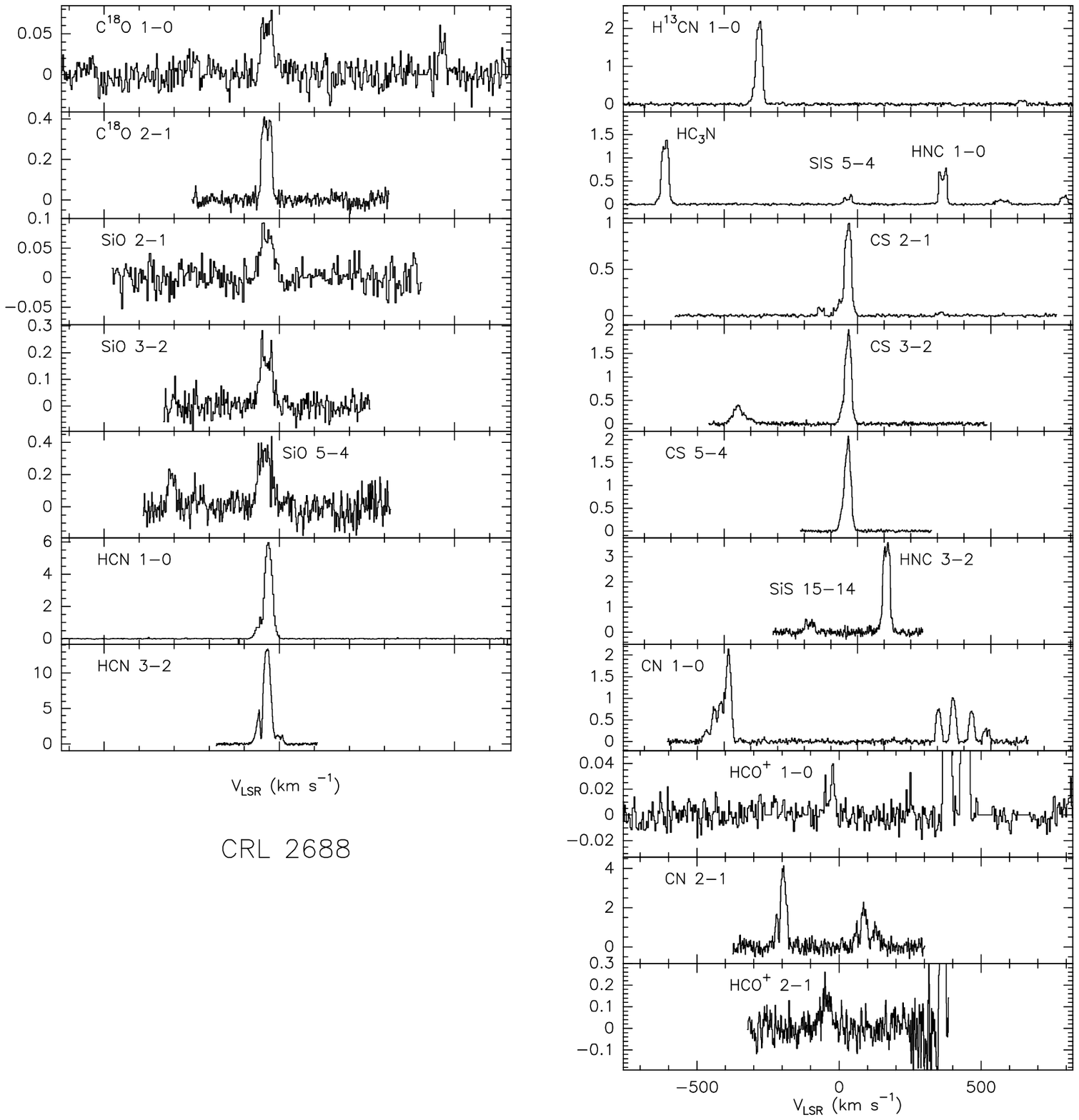}}
 \caption{Observed spectra towards CRL\,2688. The intensity scale is given in units of
 main-beam brightness temperature (K).}
\end{figure*}

\newpage

\begin{figure}
\centering
 \resizebox{\hsize}{!}{\includegraphics[angle=-90]{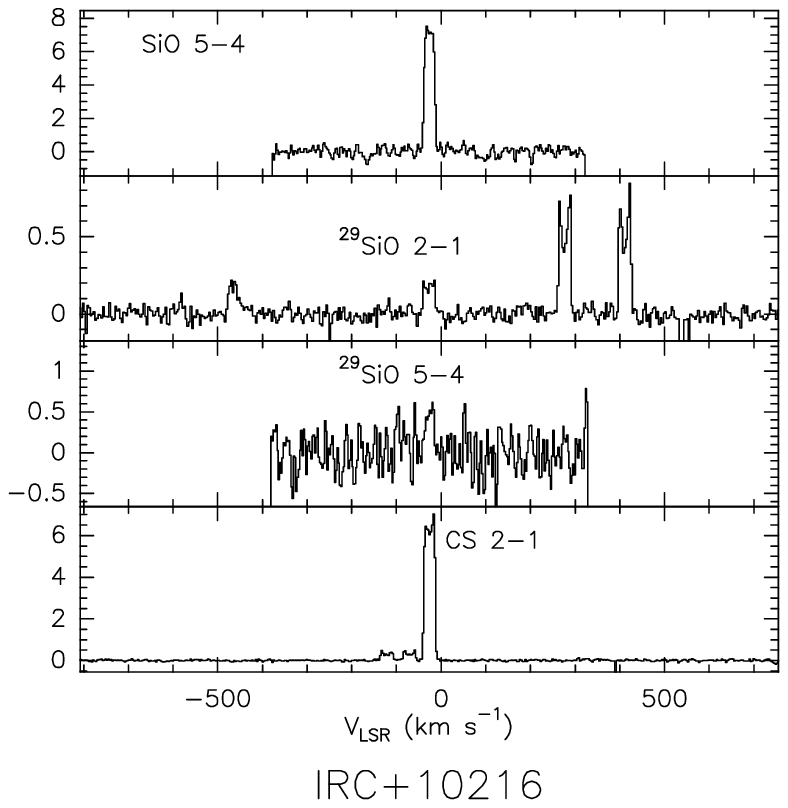}}
 \caption{Observed spectra towards IRC\,+10216. The intensity scale is given in units of
 main-beam brightness temperature (K).}
\end{figure}
\begin{figure}
\centering
 \resizebox{\hsize}{!}{\includegraphics[angle=-90]{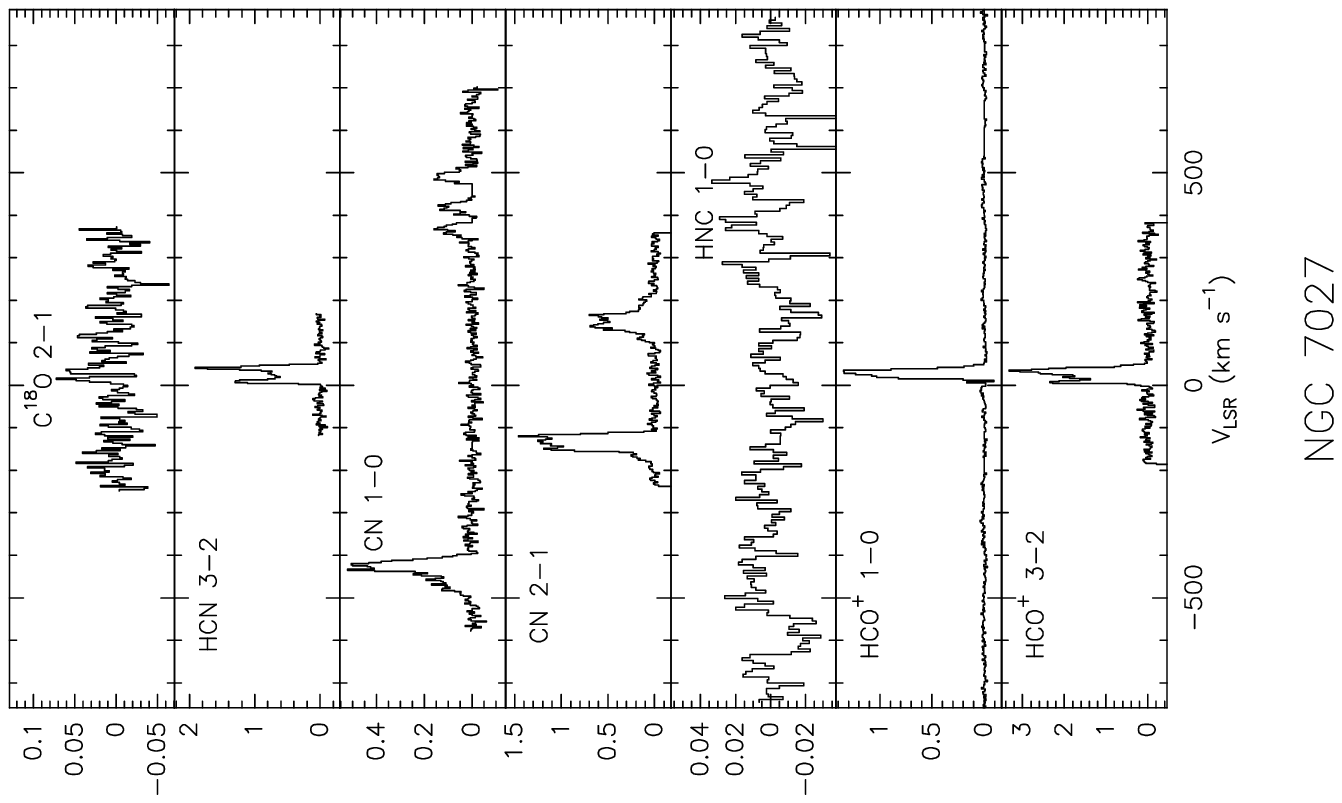}}
 \caption{Observed spectra towards NGC\,7027. The intensity scale is given in units of
 main-beam brightness temperature (K).}
\end{figure}
\begin{figure}
\centering
 \resizebox{\hsize}{!}{\includegraphics[angle=-90]{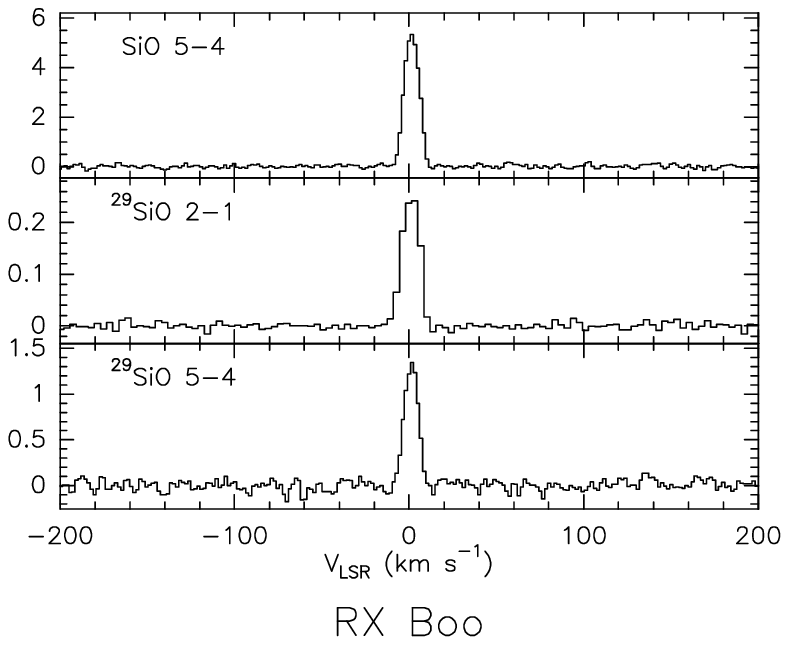}}
 \caption{Observed spectra towards RX\,Boo. The intensity scale is given in units of
 main-beam brightness temperature (K).}
\end{figure}

\begin{figure}
\centering
 \resizebox{\hsize}{!}{\includegraphics[angle=-90]{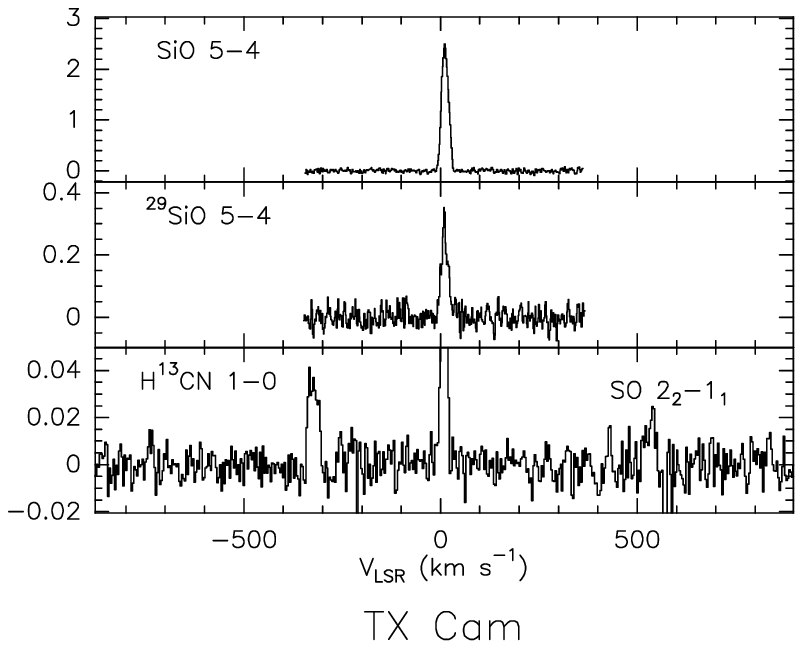}}
 \caption{Observed spectra towards TX\,Cam. The intensity scale is given in units of
 main-beam brightness temperature (K).}
\end{figure}


\end{document}